\documentstyle[12pt,aaspp4]{article} 

\lefthead{Levine \& Sparke}
\righthead{Perfect Elliptic Disks}
\slugcomment{\small Submitted to {\sl ApJ\/}, 16 Dec 1997, 
accepted 17 Mar 1998} 

\begin{document}

\title{The Stability of Perfect Elliptic Disks. \\
II. A Minimal Streaming Case}
\author{Stephen E. Levine}
\affil{Observatorio Astron\'omico Nacional, IA-UNAM \\
Ensenada, B.C., M\'exico \\ 
and \\
U. S. Naval Observatory, Flagstaff Station \altaffilmark{1} \\
P.O. Box 1149, Flagstaff, AZ 86002-1149, USA}
\authoraddr{U. S. Naval Observatory, Flagstaff Station, P.O. Box 1149
Flagstaff, AZ 86002-1149}
\and
\author{Linda S. Sparke}
\affil{Washburn Observatory, 475 North Charter Street \\ 
Madison WI 53706-1582, USA}

\altaffiltext{1}{Current address}

\begin{abstract}
Two dimensional realizations of self-consistent models for the
``perfect elliptic disks'' were tested for global stability by
gravitational N-body integration.  The family of perfect elliptic disk
potentials have two isolating integrals; time independent distribution
functions $f(E,I_2)$ which self-consistently reproduce the density
distribution can be found numerically, using a modified marching
scheme to compute the relative contributions of each member in a
library of orbits.  The possible solutions are not unique: for a given
ellipticity, the models can have a range of angular momenta.  Here
results are presented for cases with minimal angular momentum, hence
maximal random motion.  As in previous work, N-body realizations were
constructed using a modified quiet start technique to place particles
on these orbits uniformly in action-angle space, making the initial
conditions as smooth as possible.  The most elliptical models
initially showed bending instabilities; by the end of the run they had
become slightly rounder.  The most nearly axisymmetric models tended
to become more elongated, reminiscent of the radial orbit instability
in spherical systems.  Between these extremes, there is a range of
axial ratios $0.305 \lesssim b/a \lesssim 0.570$ for which the minimum
streaming models appear to be stable.
\end{abstract}


\section{Introduction}

Recent studies of elliptical galaxies and of bulges of spiral
galaxies indicate that their figures are likely to be at least
slightly triaxial (for reviews see \cite{bin82}; \cite{dzf91};
\cite{bs93}).  Most elliptical galaxies appear to be supported at
least in part by anisotropies in the velocity distributions rather
than by rapid rotation: see, for example, the work on the dwarf
elliptical galaxies NGC 147, 185 and 205 by \cite{bpn91} and
\cite{hdmp92}.  A class of non-rotating potentials, known as the
perfect ellipsoids, has been advanced as a possible model for
elliptical galaxies (e.g.  \cite{dez85}).  In these potentials, the
mass density is stratified on concentric, similar ellipsoids, and is
non-singular in the center.  Many of the properties of these
potentials can be derived analytically; the orbits all have three
isolating integrals, and hence properties such as the time-averaged
density distribution can be computed exactly.  This simplifies the
task of finding {\it self-consistent models\/}: time-steady
phase-space distribution functions $f({\bf x}, {\bf v})$ such that the
resulting mass density generates the desired gravitational potential.
\cite{stat87} and \cite{teub87} have demonstrated that distribution
functions for the perfect ellipsoids, and the analogous
two-dimensional elliptic disks, can be constructed.  Various
sub-families of the axisymmetric perfect ellipsoids have been tested
for stability (\cite{dzs89}; \cite{ms90}; \cite{mh91}; \cite{rdz91}).

Flattened perfect ellipsoids could also be viewed as models for
galactic bars.  The only analytical bar models are Freeman's
(\cite{free66a}, \cite{free66b}, \cite{free66c}) bars, which are based
upon a rotating two dimensional harmonic oscillator potential, and the
perfect elliptic disk models, which have no figure rotation.
\cite{tdz87} showed that in the limit of the needle ($b \rightarrow
0$) the two dimensional perfect elliptic disk is neutrally stable.
Prompted by the large streaming velocities seen in barred spiral
galaxies, the stability of perfect elliptic disks with maximum angular
momentum has already been studied (\cite{slls94}).  The roundest disks
were unstable to spiral mode formation, the most elongated elliptical
models were unstable to bending modes, while the models with axial
ratio $b/a$ in the range $0.250 \lesssim b/a \lesssim 0.570$ appeared
stable.  This paper extends that previous work to the study of a set
of low angular momentum perfect elliptical disks.  The minimal angular
momentum cases allow us to study the ability of internal velocity
dispersion to support an elliptic figure, and forms a natural
complement to the earlier work as the other bound of the whole class
of perfect elliptic disks. As before, we tested for global stability
by constructing a discrete, self-consistent model, loading it into an
$N$-body integrator and allowing it to evolve.

\section{Orbits in The Perfect Elliptic Disk}

\begin{figure}
\hbox to \hsize{\hfil
\vbox to 0.7\vsize{
 \includegraphics{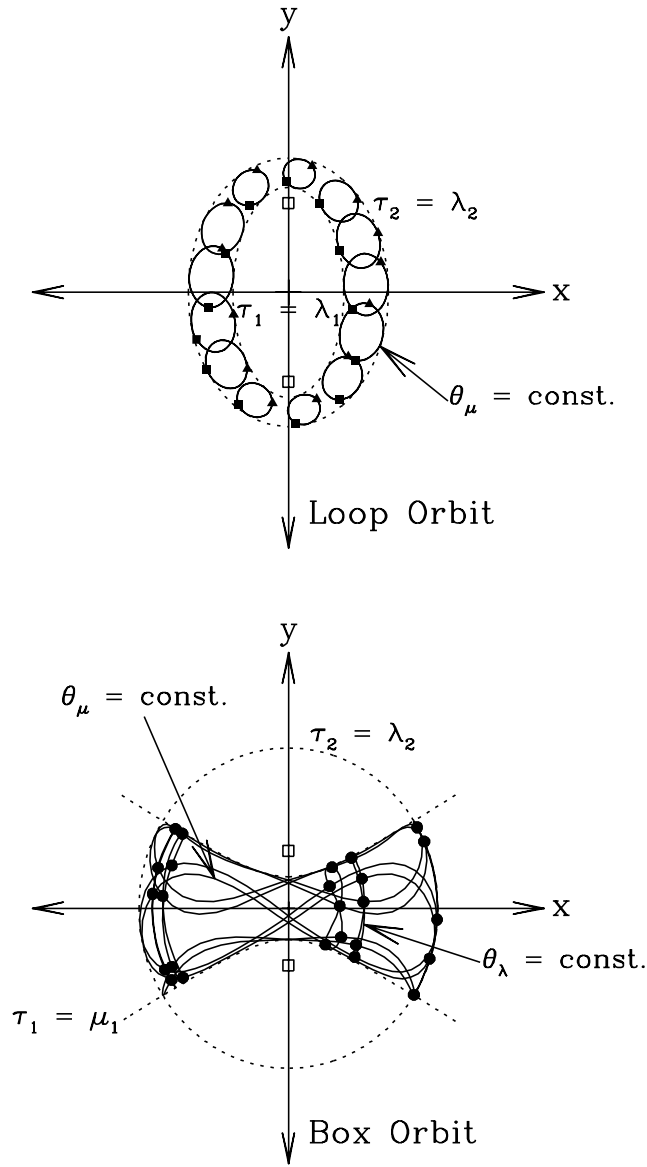}
}\hfil}
\figcaption{Sample loop (top) and box (bottom) orbits showing the
placement of particles upon each.  The orbital bounding surfaces or
turning points ($\tau_1$ and $\tau_2$) are marked with dashed lines,
and lines of constant angle have been plotted in the orbits to show
the placement of each particle at an action--angle grid intersection.
For the box orbit, the filled circles mark the intersections where
particles are placed.  For the loop orbit, for clarity, we have
plotted only lines of constant $\theta_\mu$; the intersections of
these with lines of $\theta_\lambda$ are marked with filled triangles,
and squares.  The open squares mark the foci of the elliptic
coordinates.
\label{fig-leme}}
\end{figure}

Perfect elliptic disks are the two dimensional limiting case of the
three dimensional perfect ellipsoids; their surface density $\Sigma$,
derived from the perfect ellipsoid by integrating in $z$ (see
\cite{edz92}), is given in Cartesian coordinates $(x,y)$ by
\begin{equation} \label{2ddens}
\Sigma (x,y) = {{\Sigma_0}\over{(1 + {m}^2)^{3/2}}} ~, ~~~
{m}^2 \equiv {{x^2}\over{a^2}} + {{y^2}\over{b^2}}~;
\end{equation}
$b < a$, so that $x$ is the major axis and $y$ is the minor axis.  The
elliptic disk potential satisfies St\"{a}ckel's criteria, implying
that there are two integrals of motion ($E$ and $I_2$) and that the
equations of motion are completely separable in confocal ellipsoidal
coordinates (\cite{lb62}; \cite{dez85}).  The coordinates $(\lambda,
\mu)$ are the solutions to the quadratic equation
\begin{equation} \label{coords}
{{x^2} \over {\tau + \alpha}} + {{y^2} \over {\tau + \beta}} = 1 ~,~~
\alpha < \beta < 0 ~, 
\end{equation}
where for the disk of equation (1) $-\alpha = a^2$, $-\beta = b^2$.
Curves of constant $\lambda$ are confocal ellipses aligned along the
minor axis, with $-\alpha \leq \lambda \leq \infty$, while curves of
constant $\mu$ are hyperbolae, with $-\beta \leq \mu \leq -\alpha$
(see fig. \ref{fig-leme}).  All the curves share the foci $x = 0,~ y =
\pm \sqrt{\beta-\alpha}$.  When $\lambda \approx -\alpha$, the curves
of constant $\lambda$ are highly elongated in the $y$ direction; at
large $\lambda$, they become almost circular; $\lambda +\alpha \approx
r^2$.  The curve $\mu = -\beta$ lies along the $x$-axis; as $\mu$
increases, the hyperbolae bend more sharply around the foci. 

The orbits in the elliptic disk potential divide into two families,
{\it box\/} and {\it loop\/} orbits.  The box orbits resemble
Lissajous figures, combinations of independent oscillations in the $x$
and $y$ directions; they have no net angular momentum about the center
of the potential.  Loop orbits are ellipses or rosettes with a
definite sense of rotation about the center of the potential.  Because
of the separability of the potential, the orbital bounding surfaces
(extrema of $\lambda$ and $\mu$) or {\it turning points\/} of the
orbit are all lines of constant $\lambda$ or $\mu$.  Loop orbits are
bounded by ellipses of constant $\lambda$, given by the inner and
outer turning points $\lambda_1$ and $\lambda_2$; a closed loop orbit
is a curve of constant $\lambda$.  Box orbits are contained within an
hyperbola of constant $\mu$ and an ellipse of constant $\lambda$,
corresponding to the turning points $\mu_1$ and $\lambda_2$.  Only the
loop orbits cross the minor axis outside the foci; within the foci,
only the box orbits do so (see fig. \ref{fig-leme}).  An orbit may be
characterized by its isolating integrals, by its actions, or by its
turning points.  A more complete discussion, and transformations
between the defining pairs can be found in LS, section 2.

\section{Constructing a self--consistent model}

Constructing a self-consistent model of a given potential amounts to
finding a set of orbits specified by the values $E_k$, and $I_{2k}$ of
$E$ and $I_{2}$ and a distribution weighting function $w_k$ that
together approximately reproduce the overall density distribution.
The time-average densities $\Sigma_{\rm orb}$ along these individual
orbits at any point $(\lambda, \mu)$ must sum to
\begin{equation} \label{eq-self}
\Sigma(\lambda,\mu) = \sum_{k} \Sigma_{\rm orb}(\lambda, \mu; E_k,
{I_2}_k) w_k ~~.
\end{equation}
There are two general approaches to the problem; we can either select
a set of orbits by choosing the $E_k$ and ${I_2}_k$, $k = 1, \cdots,
N$ and then compute the distribution function weights $w_k$ for each
orbit, or we can choose the weights $w_k$, and then try to find a
compatible set of orbits $E_k$ and ${I_2}_k$.

In the first method, the problem is substantially better constrained;
if there are $n$ orbits, already chosen, then we need only find $n$
weights.  In the second case, we have $n$ weights, and are trying to
find $2n$ orbit specifiers.  The first approach has been used
(\cite{schw79}; Statler 1987\nocite{stat87}; Teuben
1987\nocite{teub87}; \cite{zhs}) to demonstrate that solutions to the
self-consistent problem do exist.  If we wish to construct a discrete
representation of a potential, for an $N$-body simulation, then the
second method has the advantage of allowing us to insist that each
orbit contain an integral number of equal mass particles.  LS showed
that such problems could be solved successfully using the second
approach; here we show that the first method can also be adapted for
this purpose.  Qualitative agreement between the two methods for
the maximum angular momentum case has been shown in \cite{lev95}.

Along the minor axis, outside of the foci, the overall mass density is
comprised solely of loop orbits; this provides a constraint that
depends only on the selection of loop orbits.  Loop orbits are also
the only orbits with net angular momentum.  This allows us to split
the problem of choosing a set of orbits into two smaller problems.
First, we select a set of loop orbits which add up to give the correct
density in this region and also meet some additional criterion on the
total angular momentum, and then we find a set of box orbits which
contribute the rest of the mass needed.

\subsection{Loop Orbit Selection}

If the outer turning point $(\lambda_2)$ of a loop orbit is fixed,
then as we move the inner turning point $(\lambda_1)$ outwards, the
angular momentum of the orbit increases from a limit of zero for the
marginal orbits ($\lambda_1 = -\alpha$) up to a maximum which is
reached at the closed loop orbit $\lambda_1 = \lambda_2$.  Similarly,
for a fixed value of $\lambda_1$, the angular momentum of the orbit
increases as we increase the value of $\lambda_2$.  So, for maximum
angular momentum, we choose the thinnest possible orbits ($\lambda_1 =
\lambda_2$), and for the minimal angular momentum, the thickest
possible ($\lambda_1 \ll \lambda_2$).

In the maximum angular momentum case, the loop orbit population is
easily and uniquely determined (ZHS; LS).  The closed loops can
never overlap each other, so at any point on the minor axis outside the
foci, one and only one closed loop orbit contributes to the density at
that point; finding the mass on each of the loop orbits is a one
dimensional problem.  In LS, we chose a set of $n$ equally weighted
orbits by integrating out from the minor axis; when the total
integrated loop mass reached $1/n$ of the total loop orbit mass, we
placed a single closed orbit on the mass weighted center ellipse of
the elliptic annulus thus defined.  The outer bound of the first orbit
became the inner bound of the region represented by the next orbit,
and the process was repeated until $n$ orbits had been selected.

Any other model must contain some thick loops, and very likely some of
them will overlap each other, as well as the box orbits; finding a
satisfactory set of loop orbits has become a two dimensional problem.
We begin by dividing the minor axis into $N$ one dimensional cells,
the inner-most cell boundary being the focal point ($\lambda =
-\alpha$), and the outer-most being the outer limit in $\lambda$
$(\lambda_{\rm out})$.  We create a library of $N(N+1)/2$ orbits whose
inner and outer turning points lie on the boundaries of the cells
$(\lambda_{\rm cd})$.  For each outer turning point at $\lambda_{\rm
cd}(i=k) ~(k=1,\cdots,N)$ there are $k$ inner turning points from
$\lambda_{\rm cd}(i=0)$ (close to the marginal orbit) out to
$\lambda_{\rm cd} (i=k-1)$ (close to the closed loop orbit).  From
this library, we choose at most $N$ orbits having non-zero weight, and
compute weights $w_{\rm cd}$ for each member of the set of loop
orbits.  Because we must match the density in $N$ cells, and have
$N(N+1)/2$ possible orbits, the problem is under-constrained.  The
maximal and minimal angular momentum solutions are two cases where an
angular momentum criterion coupled with the density matching restricts
the possible solutions to a unique solution for a given library of
orbits.

To find a minimal angular momentum solution, we compute the weights
sequentially, starting with the loops with their outer turning point
in the outer-most cell.  We take the loop closest to the marginal
orbit (the ``thickest'' loop), and choose its weight so that it
accounts for all of the mass in that cell.  This fills the outermost
cell with material from the orbit of lowest angular momentum which
reachs that cell.  The mass which must still be placed in all the
inner cells is then reduced by subtracting off the contribution of
this orbit, while making sure that the density left in each cell is
non-negative.  If this is so, we move to the next outer-most cell, and
repeat the process, and so step our way in towards the center.  As we
near the center, the contributions of the low angular momentum loops
to the inner cells may fill some of them completely, before filling
the outer-most loop.  When this happens, we reduce the weight of the
outermost orbit so that the remaining mass in each cell is
non-negative, and then move the inner turning point out by one cell,
and continue the process.  We found that the loop orbit portion of the
self-consistent model cannot be made up solely of marginal orbits; it
is not possible to construct a model with exactly zero angular
momentum.

We have not proven formally that we have the minimum angular momentum
case, but empirically we appear to be close to the minumum; the
angular momenta of these models are $\sim 10^{-3}$ times less than that of
the maximum streaming models.  As a check, we tried breaking up the
solution space into an inner and an outer region at a $\lambda$ cell
boundary, and solving first the outer and then the inner regions
separately and adding up the solutions; no matter what boundary was
chosen, the angular momentum was greater than in the original single
region solution.  In addition, the derivative of the angular momentum
with respect to the inner turning point is quite steep, implying that
we really want to push towards most marginal, as we have done.

If the integrated linear density in the cells varies too greatly, we
may not be able to find a positive solution.  We separated the cell
boundaries by uniform increments in their square roots, so that the
cell divisions lie at
\begin{equation}
\label{eq-cdlcd}
\lambda_{\rm cd}(i) = \left\{ \sqrt{-\alpha + \delta} +
(\sqrt{\lambda_{\rm out}} - \sqrt{-\alpha + \delta}) \frac{i}{N}
\right\}^{2} \mbox{for $i = (0, 1, \cdots, N)$.}
\end{equation}
(The small quantity $\delta = 5 \times 10^{-5}$ is added to the
inner-most boundary to avoid the non-integrable singularity at
$\lambda = -\alpha$.)  The choice of square root spacing spreads the
distribution function weights fairly evenly among the orbits.  The
problem can be solved with other spacings (we have used linear and
exponential spacings as well though using an exponential cell spacing
for the less centrally concentrated models did not always lead to a
positive solution), but with much greater contrast in the orbit
weights.

\begin{figure}
\hbox to \hsize{\hfil \vbox to 0.7\vsize{ \includegraphics{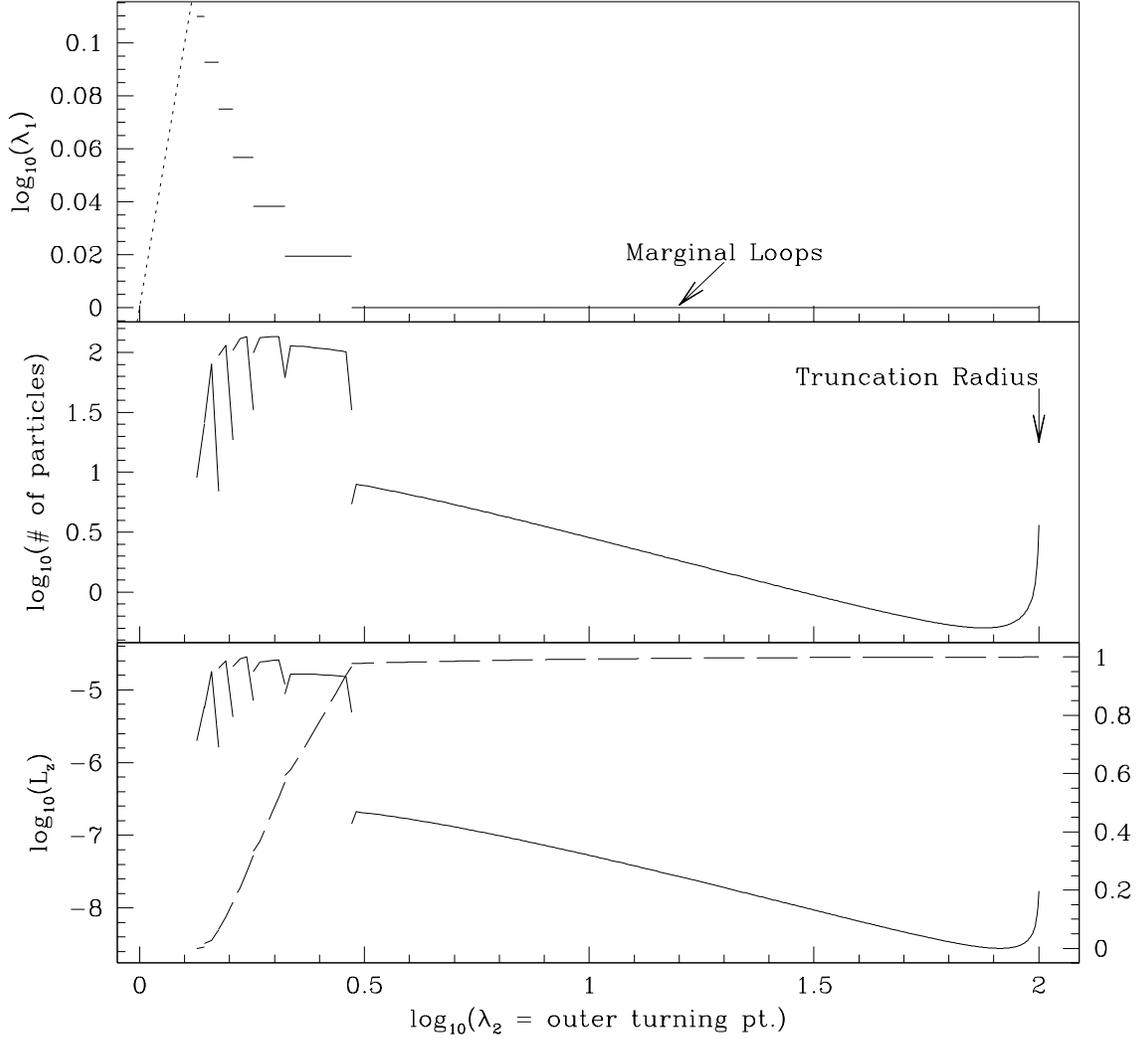} }\hfil}
\figcaption{For a model with axis ratio $b/a = 0.715$, the top panel
shows the turning points of the loop orbits with non-zero weight in
the solution.  The dashed line on the left is the line $\lambda_1 =
\lambda_2$.  The orbits with the outermost turning points are as close
to the marginal orbit as the library contains.  The middle panel shows
the number of particles assigned to each orbit as a function of the
outer turning point ($\lambda_2$).  The 400 loop orbits contain a
total of 3760 particles.  The bottom panel show the angular momentum
on each orbit (solid line) and the cumulative angular momentum as a
fraction of the total as a function of the orbit outer turning point
(dashed line).  The bulk of the angular momentum comes from the inner,
non-marginal orbits.
\label{fig-orbw}}
\end{figure}

\begin{figure}
\hbox to \hsize{\hfil \vbox to 0.7\vsize{ \includegraphics{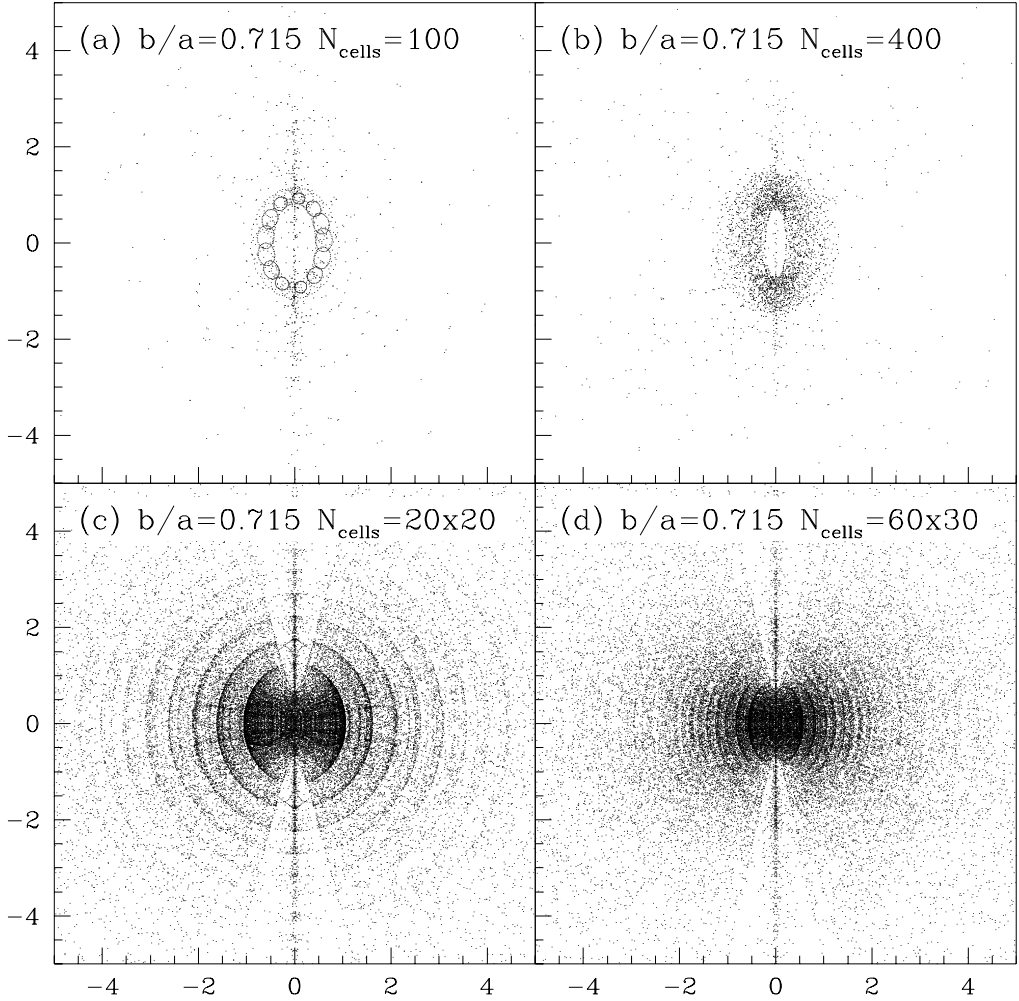} }\hfil}
\figcaption{For the model of figure 2, particle distributions on the loop
orbits for libraries with (a) 100 and (b) 400 orbits, and on the box
orbits for libraries with $N$ divisions in $\lambda$ and $M$ divisions
in $\mu$, where $N \times M$ are $20\times20$ (c) and $60\times30$ (d)
orbits respectively.  The small circles seen in (a) are a result of
using too few orbits, with the result that some of the loop orbits
have many more particles than the rest.  The solution in (a) has 1814
particles, and that in (b) has 3760.  \label{fig-numorb}}
\end{figure}

As the axial ratio gets smaller, the gradient of the density along the
minor axis (and overall) increases, making it desirable to use more
cells to achieve a better approximation of the loop orbit density.  If
the number of cells on the minor axis is too small, then the discrete
solution does not well approximate the continuous reality.
Conversely, if the number is very large, then each orbit will end up
being represented by a very small number of particles, and hence will
itself not be well sampled.  For $b/a = 0.125$, $0.570$ and $0.910$,
we computed solutions using between 100 and 1200 cells.  We then chose
the number of cells to be as small as possible and still give a good
approximation to what appeared to be the limiting continuous solution.
Figures \ref{fig-orbw} and \ref{fig-numorb} show respectively the
relative weights of orbits as a function of turning points and the
distribution of particles for 100 and 400 orbit library solutions.  We
decided to use 400 cells along the minor axis for all of the models;
this provided a good compromise between good sampling of the linear
density, and allowing high enough weights to permit us to sample each
orbit reasonably well.

\subsection{Box Orbit Selection}

\begin{figure}
 \hbox{\hfil \vbox to 2.1truein{
  \includegraphics{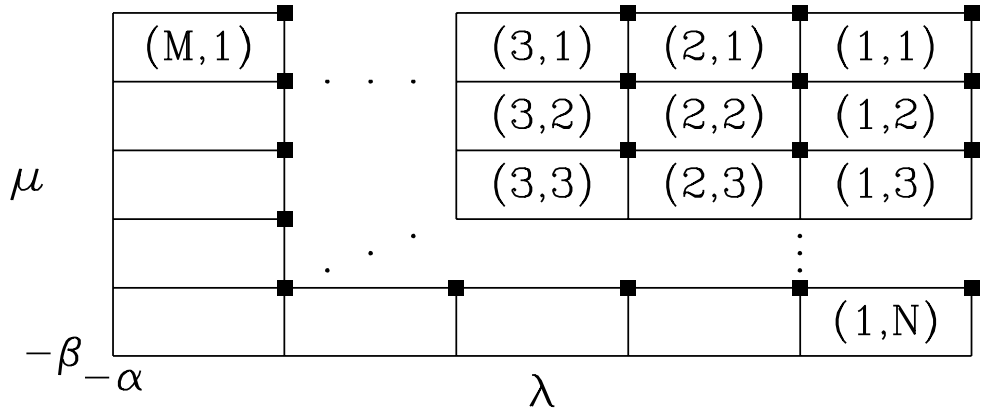}
}\hfil}
\figcaption{The ordering of the grid cells for the box orbit library, and
their respective orbits in the elliptical space in which the equations
of motion separate.  Nine of the outermost cells and orbits are shown
explicitly, with the orbit turning points labeled by filled squares.
$([x=M,\cdots,1], 1)$ denote the marginal (thickest) box orbits.
\label{fig-zhs}}
\end{figure}

For the box orbits, the method of solution follows the method of ZHS;
it is conceptually the same as for the loops, except that the grid is
a full rectangle.  First, the area of the disk is divided up into a
grid in the coordinates $(\lambda, \mu)$, with the grid points in both
$\lambda$ and $\mu$ spaced by uniform increments in the square root
(just as in the loop orbit case).  For each grid cell, the coordinates
of the corner with maximum $\lambda$ and $\mu$ are the turning points
of the box orbits in the library.  In each cell we compute the
integrated surface density minus the sum of the surface densities
already allocated to the loop orbits.  For the box orbit with turning
points in the outermost grid cell denoted $(1,1)$ in
Fig. \ref{fig-zhs}, the orbit's weight is chosen so as to supply all
the mass in the cell.  The contribution of this orbit is then
subtracted from each of the inner cells, and the process repeated for
the next outermost orbit and cell (denoted $(1,2)$), working inwards
until weights have been computed for all the orbits.  ZHS have
previously shown that positive definite solutions do exist for the
maximum streaming case; our experience is that solutions also exist
for these minimum streaming models.

We have chosen to use a library of orbits with 60 turning points
in $\lambda$ and 30 in $\mu$, making for 1800 orbits and cells.
Figure \ref{fig-numorb} shows the difference in relative sampling
between a library of $20\times20$ orbits and one with $60\times30$
orbits once they are populated with particles.

\section{Populating the Orbits}

We then place particles upon each orbit in a {\it quiet\/} manner, so
as to minimize random noise in the initial conditions, and make it
easier to watch for the growth of instabilities.  The quiet
distribution is our best approximation to a uniform distribution of
particles throughout the phase space explored by an orbit.  For a
closed orbit, the solution is easily found (\cite{sel83}).  For our
space filling orbits in an integrable potential, we use a slightly
modified version of the technique described in detail in LS.  We place
particles at the intersection points of a grid on the torus in
action--angle space which corresponds to each orbit. To find the
starting positions and velocities of each particle, we construct a
differential map from action--angle space to position and
velocity coordinates.

The weight of each orbit determines how many particles should be
placed on the it.  In all cases, we attempt to factor the integer
nearest to this number into two factors $n$, $m$ as nearly equal as
possible.  If the nearest integer is less than 6, then we accept
whatever pair of factors we compute.  Otherwise, we also factor the
second nearest integer, and choose the pair of factors that are most
nearly equal; this helps to avoid problems when the nearest integer is
prime, or has only two, very different, prime factors.  We then use a
grid with $n$ lines evenly spaced in the angle $\theta_\lambda$ and
$m$ lines in $\theta_\mu$.  The overall difference between the desired
and the computed number of particles is $\lesssim 100$ for models with
50,000 particles ($\lesssim 0.2\%$ difference), and the model produces
a good approximation of the overall potential.

\section{Model Setup and Integration}

All units from here on are expressed in terms where the gravitational
constant $G$, the length scale $a$, and the total mass $M_{\rm total}$
of each model when integrated analytically out to infinity are all
equal to 1.  Because the perfect elliptic disks are formally infinite
in extent, we have truncated the models, at $r = 10a$, for which all
the models contain at least 90\% of the total mass.  Models were
constructed for ellipticities ranging from $b/a = 0.125$ to $0.910$
following the prescription given above.  These particle distributions
were then loaded into an $N$-body integrator and allowed to evolve
under the influence of their own self-gravity.  The fraction of the
total mass within the truncation radius ($M_{\rm trunc}$), the axial
ratio ($b/a$), the relative fraction of the model mass in the loop
orbits, the angular momentum and the stability result for both the
minimum and maximum cases of each of the models are given in Table
\ref{tb-models}.

We used a two dimensional polar-grid Fourier-transform $N$-body code
developed and kindly supplied by J.  Sellwood (see \cite{sel81}, 1983
and \cite{mil76} for a more complete description).  The grid was made
up of 86 rings logarithmically spaced in radius with 100 grid points
evenly spaced about them in azimuth.  The grid was bounded by circles
of radius $r = 0.05a$ and $r = 10.4a$.  The outermost cells had
dimensions $\Delta r = 0.64a$ by $r \Delta \theta = 0.66a$.  Stars
that crossed the outer boundary during the integration were discarded,
while stars crossing into the central hole continued across it with
constant velocity on a straight path, and were placed back on the grid
at the next step (\cite{sel83}).  An explicit softening length of
$0.05$ was used for these integrations, and a small compensating
radial force correction was added to insure that our initial models
were close to virial equilibrium.  For an extended discussion of the
details of the integration, and determination of the softening length
see LS.

The integration time step $\Delta t$ was chosen to be less than
$1/20^{\rm th}$ of the minimum period required for each of the angle
variables to complete a circuit of $2\pi$ radians on the action--angle
torus for at least 90\% of the orbits.  We used $\Delta t = 0.01$ for
all the models except $b/a = 0.125$ and $0.180$, for which $\Delta t =
0.002$; the most elliptic models have higher central densities and
velocities, requiring better time resolution.  All of the models were
run for at least 20 dimensionless time units, corresponding to 3--6
crossing times at the half-mass radius.  This was enough time for
gross instabilities to develop.  For a number of the models we continued
the simulations for 100 times units (15--30 half-mass crossing times).

At the beginning of the runs, the number of particles in the central
hole was less than $0.5{\rm \%}$.  In the most unstable models,
several hundred particles fell off the grid within $t=20$, while the
stable models lost only tens of particles out of 50,000.  The more
elliptic models lost more particles, perhaps because of the higher
radial velocities in their deeper central potentials.  By $t=100$,
between $2{\rm \%}$ and $14{\rm \%}$ of the particles had left the
grid.

\section{Results}

\begin{figure}
{\hfuzz=300pt
 \hbox to \hsize{\hskip 1truecm
 \vbox to 0.88\vsize{
  \includegraphics{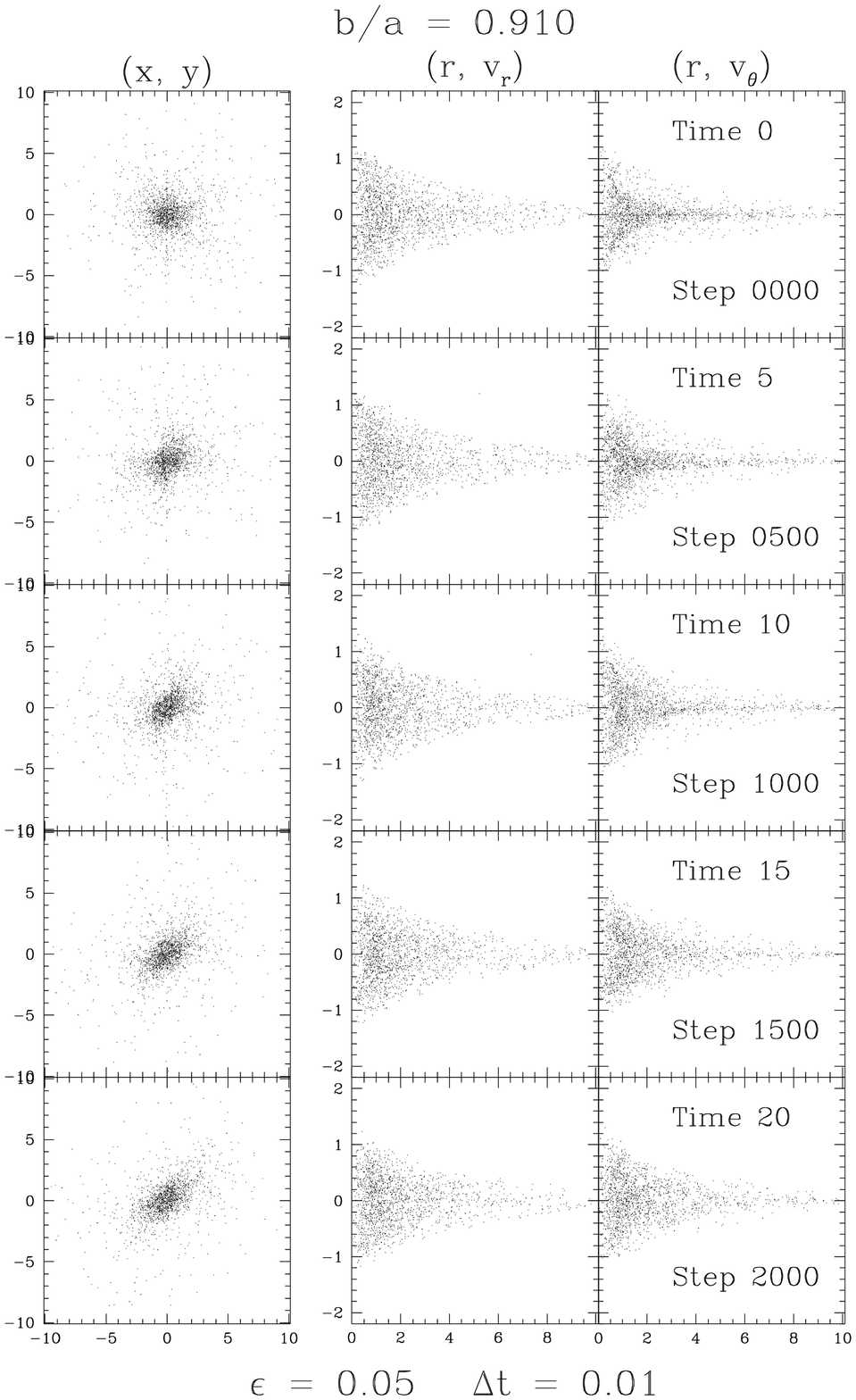}
 }}
}
\figcaption{Positions ({\it left column\/}) and velocities (radial
velocity: {\it middle column\/}; tangential velocity: {\it right
column\/}) for 2500 of the 50,000 particles in the $N$-body integrations
are shown at times 0, 5, 10, 15 and 20, for $b/a = 0.910$. 
\label{fig-run910}}
\end{figure}

\begin{figure}
{\hfuzz=300pt
 \hbox to \hsize{\hskip 1truecm
 \vbox to 0.88\vsize{
  \includegraphics{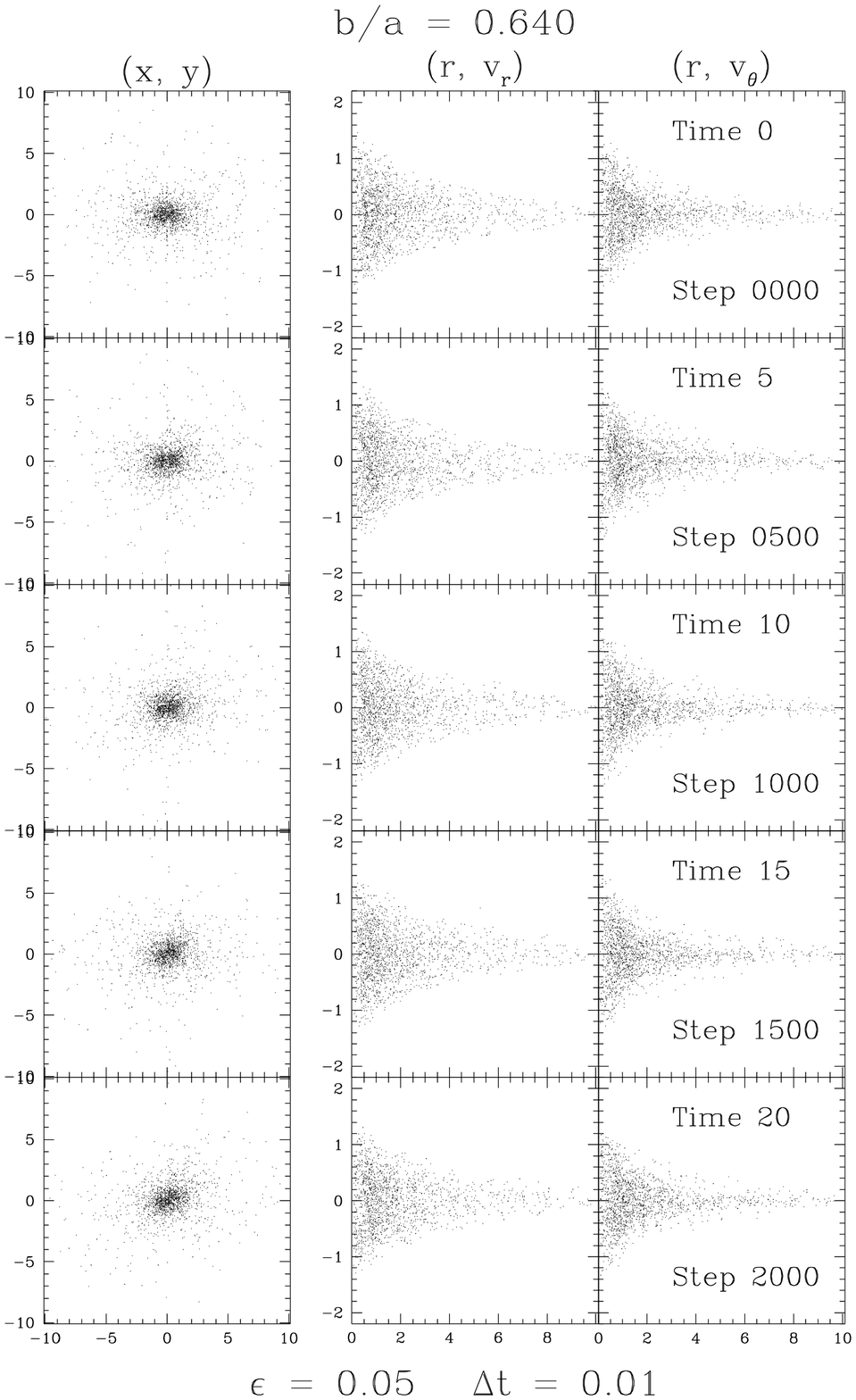}
 }}
} 
\figcaption{Positions ({\it left column\/}) and velocities (radial
velocity: {\it middle column\/}; tangential velocity: {\it right
column\/}) for 2500 of the 50,000 particles in the $N$-body integrations
are shown at times 0, 5, 10, 15 and 20, for $b/a = 0.640$. 
\label{fig-run640}}
\end{figure}

\begin{figure}
{\hfuzz=300pt
 \hbox to \hsize{\hskip 1truecm
 \vbox to 0.88\vsize{
  \includegraphics{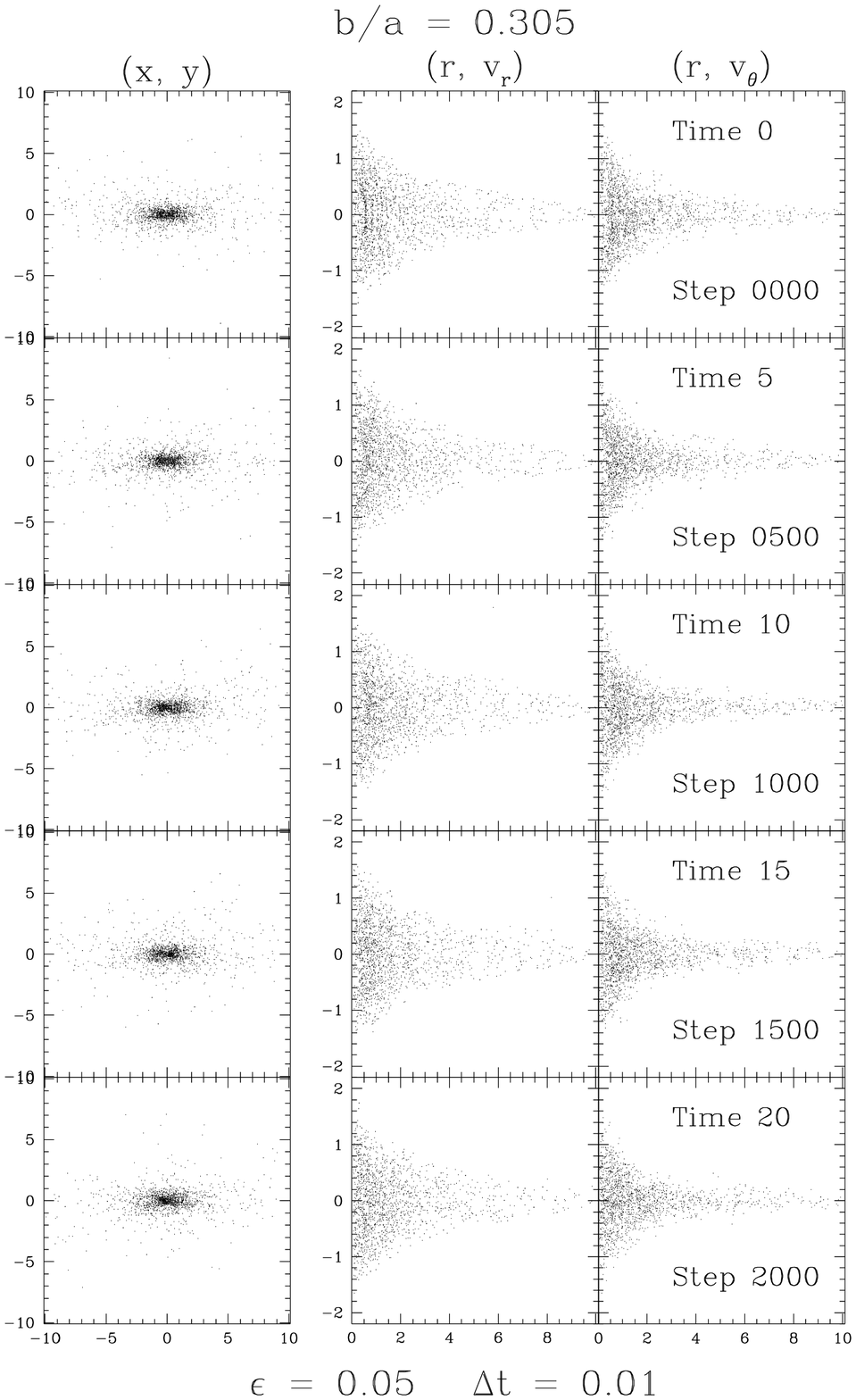}
 }}
} 
\figcaption{Positions ({\it left column\/}) and velocities (radial
velocity: {\it middle column\/}; tangential velocity: {\it right
column\/}) for 2500 of the 50,000 particles in the $N$-body integrations
are shown at times 0, 5, 10, 15 and 20, for $b/a = 0.305$. 
\label{fig-run305}}
\end{figure}

\begin{figure}
{\hfuzz=300pt
 \hbox to \hsize{\hskip 1truecm
 \vbox to 0.88\vsize{
  \includegraphics{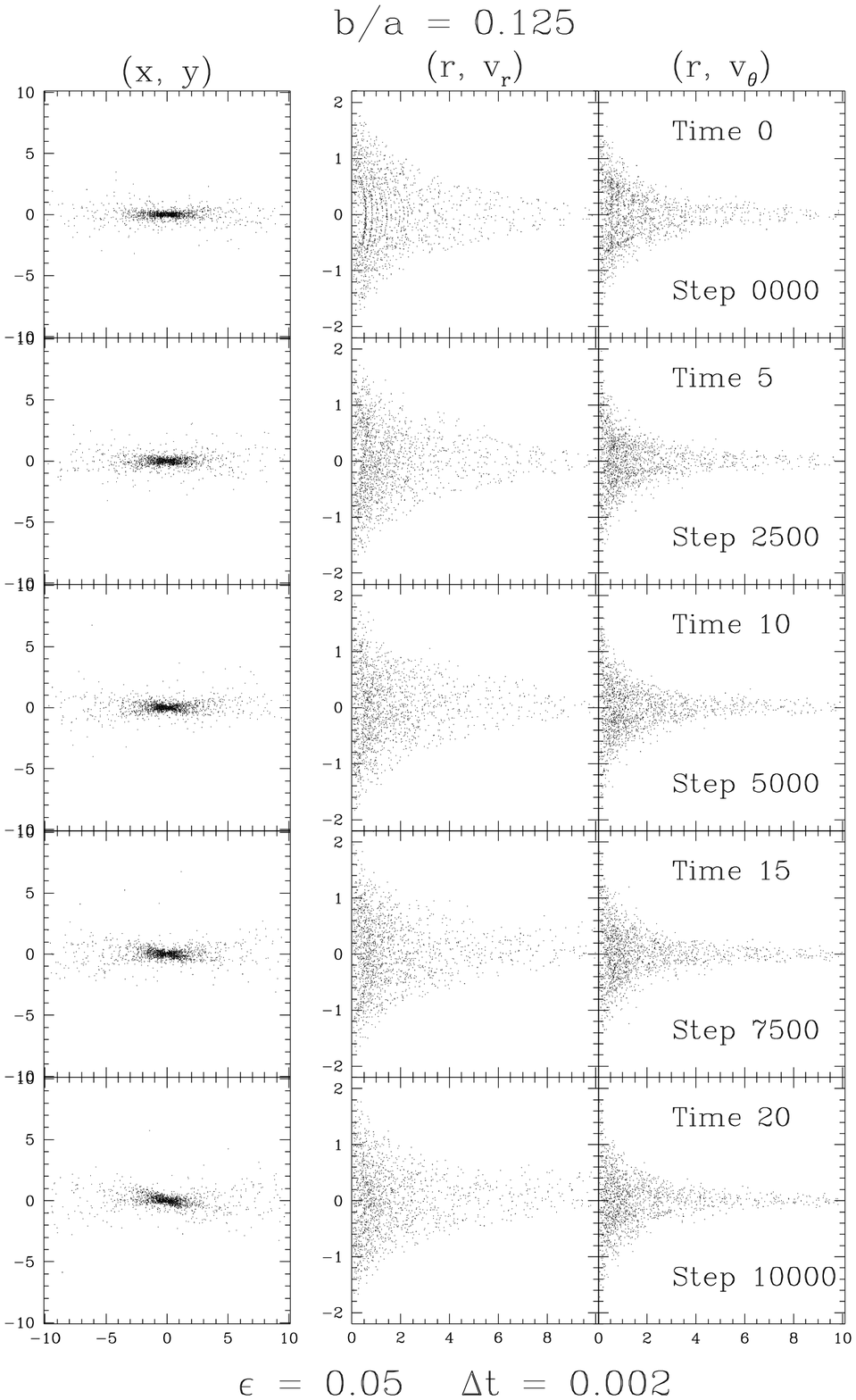}
 }}
} 
\figcaption{Positions ({\it left column\/}) and velocities (radial
velocity: {\it middle column\/}; tangential velocity: {\it right
column\/}) for 2500 of the 50,000 particles in the $N$-body integrations
are shown at times 0, 5, 10, 15 and 20, for $b/a = 0.125$. 
\label{fig-run125}}
\end{figure}

We constructed a set of minimal angular momentum models with
ellipticities ranging from $0.125$ to $0.910$, highly elongated to
almost circular.  The minimal angular momentum models at both extremes
of axial ratio appear to be unstable.  For the nearly axisymmetric
models (for $b/a = 0.910$ to $0.640$, see fig. \ref{fig-run910} and
\ref{fig-run640}), the dominant instability manifests itself as an
increase in ellipticity, resembling the radial orbit instability seen
in three dimensional simulations that have little rotational support
(e.g. \cite{ma85}; \cite{bgh86}).  In the most elongated models
(fig. \ref{fig-run125}) where $b/a \lesssim 0.2$, we see initially
the beginnings of a bending instability, just as in the maximal
streaming models, and similar to the bending seen in the prolate E9
model of Merritt \& Hernquist (1991)\nocite{mh91}.  This is followed
later by a decrease in the overall ellipticity of the figure, and a
growth in the power in the ``lopsided'' $m=1$ mode similar to that
seen in \cite{pp90} and \cite{ppa90}.  Models with axial ratio ratio
between $0.305$ and $0.570$ appear to be stable over the duration of
the simulations (fig. \ref{fig-run305}).

\begin{figure}
{\hfuzz=300pt
 \hbox to \hsize{\hskip 1truecm
 \vbox to 0.8\vsize{
  \includegraphics{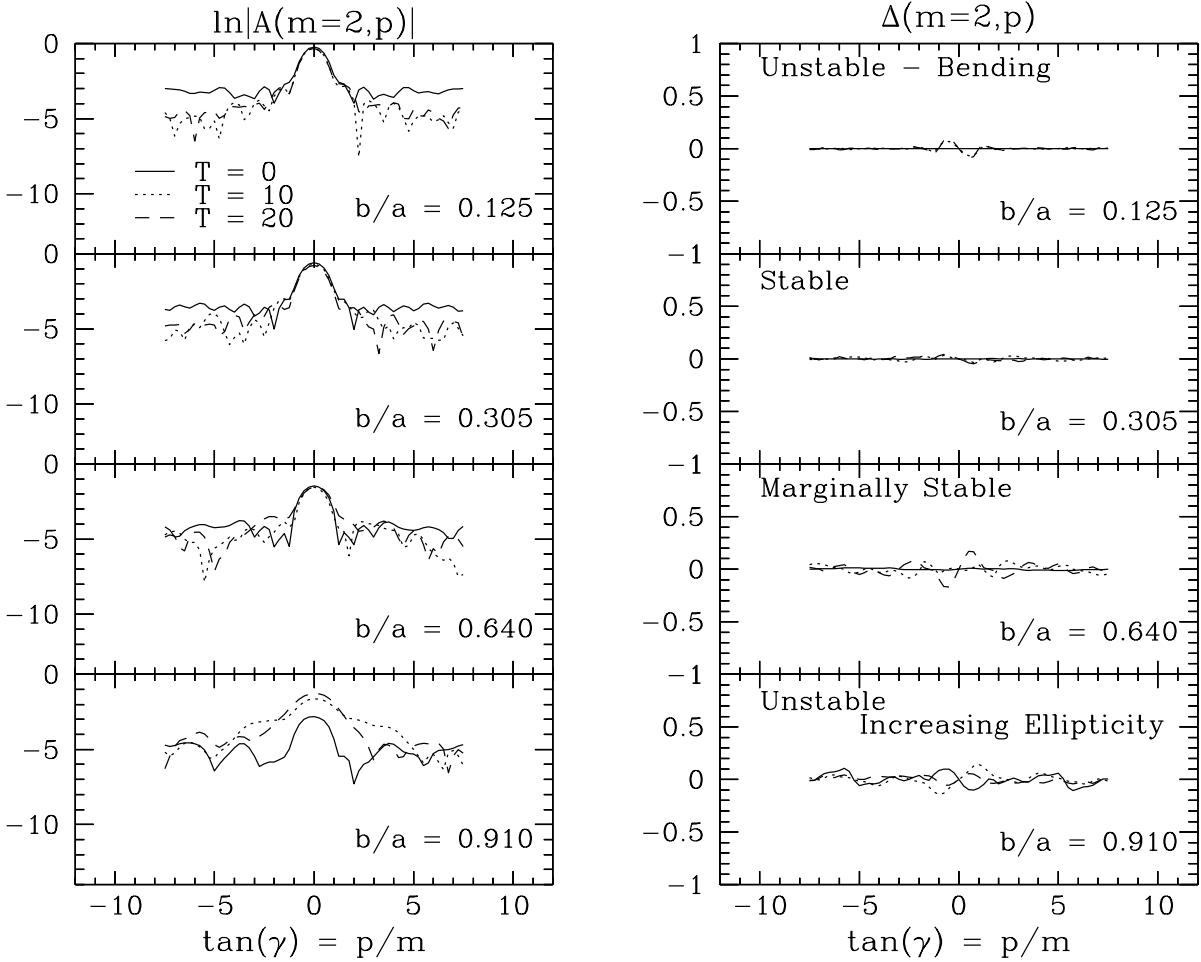}

 }}
}
\figcaption{The left hand column shows the distribution of power with
the tangent of the pitch angle in the $m=2$ log spiral mode ($\ln
|A(m=2,p)|$) for models with $b/a = $ $0.125$, $0.305$, $0.640$, and
$0.910$.  The right hand column shows the growth of the asymmetry term
$\Delta (m=2,p)$ in the $m=2$ mode.  These are plotted at times 0
({\it solid line\/}), 10 ({\it dotted line\/}) and 20 ({\it dashed
line\/}). \label{fig-amp}}
\end{figure}

Since the growth of an instability is not always apparent in the plots
of particle position, we have also examined the behavior of the first
six logarithmic spiral coefficients for these models.  The growth of
asymmetry in the $m=2$ mode was defined in LS eq. [31] as
\begin{equation} \label{lgsym}
\Delta (m=2,p) = {{|A(m=2,p)| -
|A(m=2,-p)|} \over {|A(m=2,p=0)|}} ~~;
\end{equation}
this measures the spirality which is inherent in growing modes
(\cite{lo67}; LS, eq. [27] \& [31]).  Incipient spiral instabilities
can show up here before they are clearly visible in the plots of
particle positions.  We call ``stable'' those models for which there
is no change, above the noise level inherent in the particle
discreteness, in the amplitude of the spiral harmonics.  Figure
\ref{fig-amp} shows the change over the course of an $N$-body
integration of the dominant $m=2$ log spiral mode, for the models of
figures \ref{fig-run910}--\ref{fig-run125}.  Bending-unstable models
such as $b/a = 0.125$ grow asymmetrically at small $p/m$; this is most
apparent as the s-shaped curve in the symmetry plot.  The unstable
nearly-round models which become more elliptic show a mostly symmetric
growth in the $m=2$ power.  Stable models such as $b/a = 0.305$ don't
budge.  We have labeled the models with $b/a = 0.250$ and $0.640$ as
marginally stable: we see global instabilities of small power develop,
but these very quickly saturate and die out.

\section{Discussion}

In this paper and in LS, we have constructed discrete self-consistent
representations of the distribution functions of a range of perfect
elliptic disks with minimal and maximal angular momentum.  These
models were then integrated forward in time using an $N$-body
integrator to see if they were stable.  The nearly axisymmetric and
the most elongated models were unstable.  The perfect elliptic disks
with moderate axial ratios appear to be stable in both the maximum and
minimum streaming cases.

In the maximum streaming case, the nearly axisymmetric models
developed spiral and bar instabilities as expected, since their
limiting case, a cold axisymmetric disk, is known to be violently
unstable to spiral instabilities.  In the minimal angular momentum
case, nearly-round disks became more elliptical, in a manner very
similar to the radial orbit instability of spherical systems.  This is
not too surprising, since the velocity distribution is anisotropic,
with the radial velocity dispersion being substantially higher than
the tangential dispersion, even in the very nearly axisymmetric
models.  This comes about because of the substantial presence of box
and marginal loop orbits in the models.

In both angular momentum extremes, the most elongated models developed
a bending instability.  The similarity in behavior is not very
surprising given the decreasing importance of rotational support with
increasing ellipticity in these models.  \cite{tdz87} have shown that
the limiting case of the needle ($b\rightarrow0$) is neutrally stable
to bending, while Merritt \& Hernquist (1991)\nocite{mh91} have
demonstrated a bending instability in a very prolate (E9) system.  It
is thus not surprising that the most elliptic models should develop
this instability.  For the minimum angular momentum family, the
instabilities change the shape of the disk towards a more moderate
ellipticity.

\begin{figure}
\hbox to \hsize{\hfil
 \vbox to 0.7\vsize{
  \includegraphics{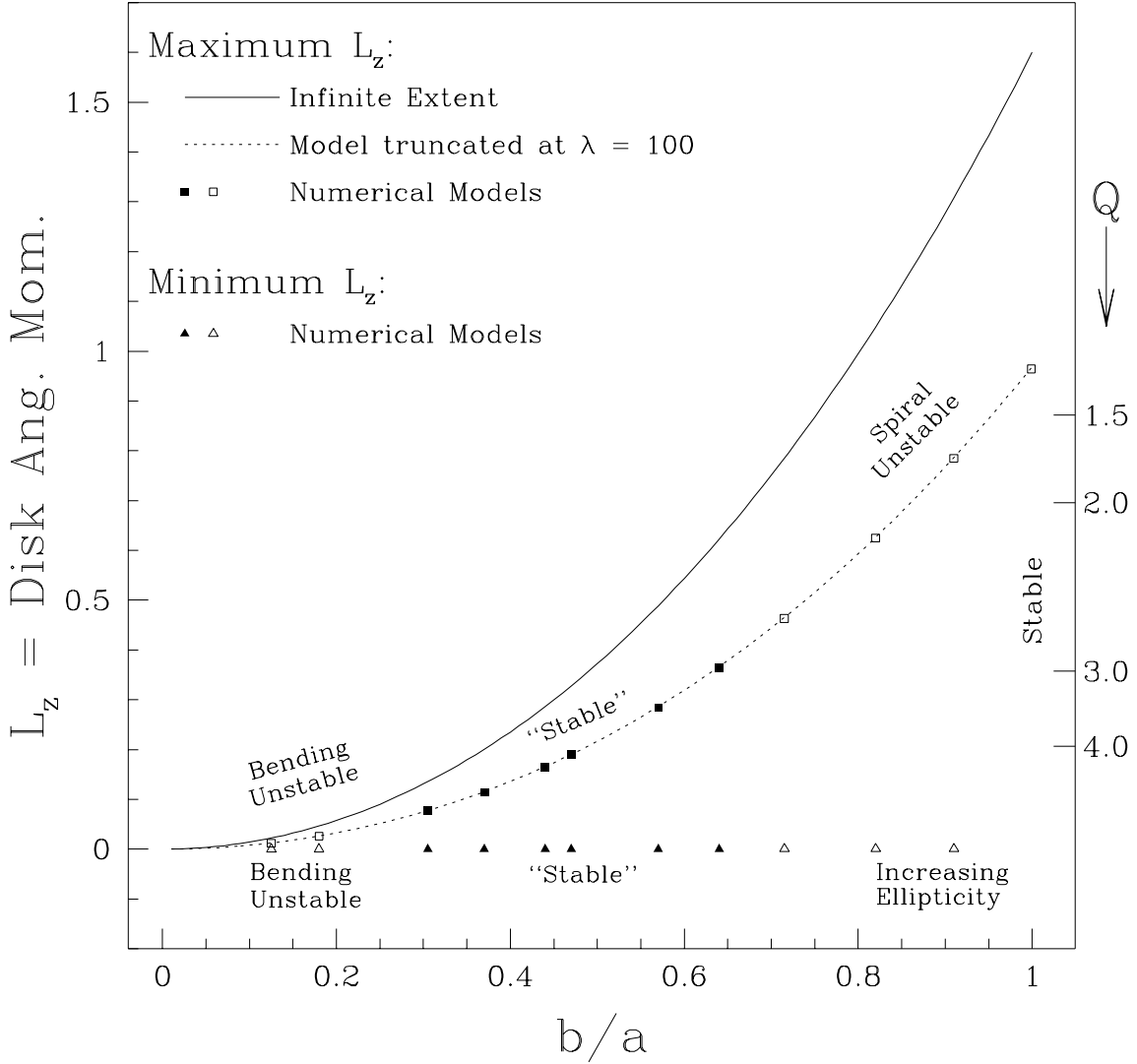}
}\hfil}
\figcaption{Disk angular momentum as a function of ellipticity for the
maximum and minimal angular momentum disks.  The solid line shows the
angular momentum in the infinite maximum angular momentum disks, the
squares the actual values in the truncated maximum angular momentum
disk models, and the triangles the values for the minimal angular
momentum disk models.  Open and filled symbols represent unstable and
stable models respectively.  Values of $Q$ for axisymmetric models are
indicated along the right hand axis.  $T_r/T_t = 0.85$ and $1.14$ for $Q
= 2$ and $3$ respectively.  \label{fig-angmom}}
\end{figure}

In the nearly axisymmetric disks, as the angular momentum is decreased
from a maximum, we expect that the increasing velocity dispersion
should help to stabilize against spiral instabilities.  It appears
likely that there is a stable region for nearly axisymmetric disks
with values of Toomre's (\cite{t64}) stability parameter $Q$ which
lies between the points $Q \sim 2$ and $Q \sim 3$ where the velocity
dispersion has increased to the point of being able to support the
disk against the spiral instability (fig \ref{fig-angmom}).  As the
rotational support becomes negligible and the radial velocity
dispersion increases, a radial--orbit instability develops; the disks
with lower angular momentum become unstable to elliptical distortions
when $T_{\rm radial} / T_{\rm tangential} \gtrsim 1.2$ (after the
discussion of \cite{fp84} and \cite{p87} for stability of spherical
systems).  We expect that there is a range of angular momentum between
the two extremes for which the nearly round disks are stable.  The
moderately elliptical disks with maximum and minimum angular momentum
appear to be stable, so we would anticipate that disks of similar
ellipticity and intermediate angular momentum will also be stable.

The stability of the two-dimensional models with moderate ellipticity
gives us hope that the three dimensional perfect ellipsoids of
intermediate triaxiality (which is probably the appropriate range for
elliptical galaxies \cite{mb81}; \cite{ddc76}; \cite{dzf91}), will
also be stable.  It is known that some very flattened systems, such as
the extreme oblate spheroids constructed from thin short-axis tube
orbits (\cite{ms90}), are unstable, but the simple fact that two
longer axes are unequal is not likely to be the cause of further
trouble.  The three dimensional extension of this work will be
interesting to see in light of the work of \cite{app92} showing that
three dimensional systems with a small amount of rotational streaming
are unstable to a tumbling bar instability, both when the models have
largely radial orbits and when the orbits are mostly circular.

The techniques developed in this work have laid the foundation for
investigating the stability of three dimensional perfect ellipsoids,
and indeed of any integrable potential.  The methods for choosing
orbits, whether simulated annealing (as in LS) or the marching scheme
of ZHS,
can be easily expanded to take account of a variety of possible
cost terms related to angular momentum or line of sight velocities
(e.g. \cite{rix97}).  The procedure of LS for
generating a quiet start which minimizes random noise due to particle
discreteness, can be carried over to any integrable system.  This is
potentially most useful in $N$-body studies which attempt to measure
the growth rate of instabilities, because the detection of
instabilities which are still in the linear regime is limited by
particle noise.  For example, \cite{s91} found that his linear
stability theory was consistent with the results of $N$-body
simulations for highly unstable spherical systems, but predicted slow
growing instabilities which could not be seen in the simulations
because of particle noise.  \cite{app90} have constructed an analytic
potential--smoothing integration technique which decreases the
$\sqrt{N}$ noise associated with binning and softening in $N$-body
codes, and permits better examination of the linear growth regime.
Their method would also benefit from a quiet start, because the
particle discreteness then makes a larger relative contribution to the
noise.

\acknowledgments The authors would like to thank P. T. de~Zeeuw for
continued interest and for suggesting that we make sure that the ZHS
and LS schemes agree, and J. Sellwood for graciously allowing us to
use his $N$-body integrator.  This work has received support from
grants 3739--E from the Consejo Nacional de Ciencia y Tecnolog\'{\i}a
of Mexico, and NAGW--2769 from the National Aeronautics and Space
Administration of the USA.

\newpage

\begin{deluxetable}{cccccccc}
\small
\tablewidth{6in}
\tablecaption{Model Details \label{tb-models}}
\tablehead{
\colhead{} & \colhead{} & \multicolumn{3}{c}{Minimum $L_z$} &
\multicolumn{3}{c}{Maximum $L_z$} \\ 
\cline{3-5} \cline{6-8} \\
\colhead{$b/a$} & 
\colhead{$M_{\rm trunc}$} & 
\colhead{$\frac{M_{\rm loop}}{M_{\rm trunc}}$} &
\colhead{${10^3 \times L_{\rm z}}$} &
\colhead{Stable\tablenotemark{a}} &
\colhead{$\frac{M_{\rm loop}}{M_{\rm trunc}}$} &
\colhead{${10^3 \times L_{\rm z}}$} &
\colhead{Stable\tablenotemark{a}}}
\startdata
0.125 & 0.935 & 0.002 & 0.012 & Bend     & 0.01 & {\phn}11.98 & Bend      \nl
0.180 & 0.934 & 0.003 & 0.025 & Bend     & 0.02 & {\phn}26.31 & Bend      \nl
0.250 & 0.932 & 0.007 & 0.051 & Marginal & 0.04 & {\phn}50.84 & Stable    \nl
0.305 & 0.930 & 0.010 & 0.078 & Stable   & 0.06 & {\phn}76.68 & Stable    \nl
0.370 & 0.928 & 0.015 & 0.120 & Stable   & 0.09 &      114.80 & Stable    \nl
0.440 & 0.926 & 0.022 & 0.177 & Stable   & 0.13 &      165.27 & Stable    \nl
0.470 & 0.924 & 0.026 & 0.205 & Stable   & 0.15 &      189.75 & Stable    \nl
0.570 & 0.920 & 0.040 & 0.312 & Stable   & 0.23 &      284.66 & Stable    \nl
0.640 & 0.917 & 0.054 & 0.399 & Marginal & 0.30 &      364.85 & Marginal  \nl
0.715 & 0.914 & 0.074 & 0.497 & Elliptic & 0.39 &      463.27 & Spiral    \nl
0.820 & 0.909 & 0.115 & 0.629 & Elliptic & 0.55 &      624.68 & Spiral    \nl
0.910 & 0.905 & 0.176 & 0.708 & Elliptic & 0.73 &      784.92 & Spiral    \nl
\enddata

\tablenotetext{a}{Bend denotes bending into an `S' shape, with
decreasing $b/a$, Elliptic means unstable to increasing ellipticity,
Spiral is unstable to forming a spiral.}
\end{deluxetable}


\newpage

\newpage

\newpage

\newpage

\newpage

\newpage

\newpage

\newpage

\newpage

\newpage

\newpage


\begin{thebibliography}{}

\bibitem[Allen, Palmer \& Papaloizou (1990)]{app90}
 Allen, A. J., Palmer, P. L., \& Papaloizou, J. C. B. 1990, \mnras, 242, 576

\bibitem[Allen, Palmer \& Papaloizou (1992)]{app92}
 Allen, A. J., Palmer, P. L., \& Papaloizou, J. C. B. 1992, \mnras, 256, 695

\bibitem[Barnes, Goodman \& Hut 1986]{bgh86}
 Barnes, J., Goodman, J., \& Hut, P. 1986, \apj, 300, 112

\bibitem[Bender, Paquet \& Nieto (1991)]{bpn91}
 Bender, R., Paquet, A., \& Nieto, J.-L. 1991, \aap, 246,349

\bibitem[Bertin \& Stiavelli 1993]{bs93}
 Bertin, G., \& Stiavelli, M. 1993, Rep. Prog. Phys., 56, 493

\bibitem[Binney 1982]{bin82}
 Binney, J. 1982, \araa, 20, 399

\bibitem[de~Vaucouleurs, de~Vaucouleurs \& Corwin 1976]{ddc76}
 de~Vaucouleurs, G., de~Vaucouleurs, A., \& Corwin, H. G. 1976, Second
  Reference Catalogue of Bright Galaxies (Austin: University of Texas Press)

\bibitem[de~Zeeuw 1985]{dez85}
 de~Zeeuw, P. T. 1985, \mnras, 216, 273

\bibitem[de~Zeeuw \& Franx 1991]{dzf91}
 de~Zeeuw, P. T., \& Franx, M. 1991, \araa, 29, 239

\bibitem[de~Zeeuw, Hunter \& Schwarzschild 1987, hereafter ZHS]{zhs}
 de~Zeeuw, P. T., Hunter, C., \& Schwarzschild, M. 1987, \apj, 317, 607 (ZHS)

\bibitem[de~Zeeuw \& Schwarzschild 1989]{dzs89}
 de~Zeeuw, P. T., \& Schwarzschild, M. 1989, \apj, 345, 84

\bibitem[Evans \& de~Zeeuw 1992]{edz92}
 Evans, N. W., \& de~Zeeuw, P. T. 1992, \mnras, 257, 152

\bibitem[1966a]{free66a}
 Freeman, K. C. 1966a, \mnras, 133, 47

\bibitem[b]{free66b}
 Freeman, K. C. 1966b, \mnras, 134, 1

\bibitem[c]{free66c}
 Freeman, K. C. 1966c, \mnras, 134, 15

\bibitem[Fridman \& Polyachenko 1984]{fp84}
 Fridman, A., \& Polyachenko, V. 1984, Physics of Gravitating Systems,
 vols. 1 and 2 (Berlin: Springer-Verlag)

\bibitem[Held et al. (1992)]{hdmp92}
 Held, E., de~Zeeuw, P. T., Mould, J., \& Picard, A. 1992, \aj, 103, 851

\bibitem[Levine (1995)]{lev95}
 Levine, S. E. 1995, in The Formation of the Milky Way, ed. E. J.  Alfaro
 \& A. J. Delgado (New York: Cambridge Univ. Press), 247

\bibitem[Levine \& Sparke 1994, LS]{slls94}
 Levine, S. E., \& Sparke, L. S. 1994, \apj, 428, 493 (LS)

\bibitem[Lynden-Bell 1962]{lb62}
 Lynden-Bell, D. 1962, \mnras, 124, 95

\bibitem[Lynden-Bell \& Ostriker 1967]{lo67}
 Lynden-Bell, D., \& Ostriker, J. P. 1967, \mnras, 136, 239

\bibitem[Merritt \& Aguilar 1985]{ma85}
 Merritt, D., \& Aguilar, L. 1985, \mnras, 217, 787

\bibitem[Merritt \& Hernquist 1991]{mh91}
 Merritt, D., \& Hernquist, L. 1991, \apj, 376, 439

\bibitem[Merritt \& Stiavelli 1990]{ms90}
 Merritt, D., \& Stiavelli, M. 1990, \apj, 358, 399

\bibitem[Mihalas \& Binney 1981]{mb81}
 Mihalas, D., \& Binney, J. 1981, Galactic Astronomy (San Francisco:
  W. H. Freeman)

\bibitem[Miller 1976]{mil76}
 Miller, R. H. 1976, J. Comp. Phys., 21, 400

\bibitem[Palmer \& Papaloizou (1990)]{pp90}
 Palmer, P. L., \& Papaloizou, J. C. B. 1990, \mnras, 243, 263

\bibitem[Palmer, Papaloizou \& Allen (1990)]{ppa90}
 Palmer, P. L., Papaloizou, J. C. B., \& Allen, A. J. 1990, \mnras, 243, 282

\bibitem[Polyachenko 1987]{p87}
 Polyachenko, V. 1987, in IAU Symp. 127, Structure and Dynamics of
 Elliptical Galaxies, ed. P. T. de~Zeeuw (Dordrecht: Reidel), 301

\bibitem[Rix 1997]{rix97}
 Rix, H.-W., de~Zeeuw, P. T., Cretton, N., van~der~Marel, R.P.,
 Carollo, C. M. 1997, \apj, 488, 702

\bibitem[Robijn \& de~Zeeuw 1991]{rdz91}
 Robijn, F., \& de~Zeeuw, P. T. 1991, in Dynamics of Disc Galaxies,
 ed. B. Sundelius (G{\"o}teborg, Sweden), 373

\bibitem[Saha (1991)]{s91}
 Saha, P. 1991, \mnras, 248, 494

\bibitem[Schwarzschild 1979]{schw79}
 Schwarzschild, M. 1979, \apj, 232, 236

\bibitem[Sellwood 1981]{sel81}
 Sellwood, J. A. 1981, \aap, 99, 362

\bibitem[Sellwood 1983]{sel83}
 Sellwood, J. A. 1983, J. Comp. Phys., 50, 337

\bibitem[Statler (1987)]{stat87}
 Statler, T. 1987, \apj, 321 113

\bibitem[Teuben (1987)]{teub87}
 Teuben, P. J. 1987, \mnras, 227, 815

\bibitem[1964]{t64}
 Toomre, A. 1964, \apj, 139, 1217

\bibitem[Tremaine \& de~Zeeuw (1987)]{tdz87}
 Tremaine, S., \& de~Zeeuw, P. T. 1987, in IAU Symp. 127,
 Structure and Dynamics of Elliptic Galaxies, ed. P. T. de~Zeeuw
 (Dordrecht: Reidel), 493

\end{thebibliography}
\end{document}